\documentclass[rmp,floatfix,preprint,endfloats*]{revtex4}

\usepackage{natbib}

\usepackage{times,setspace}
\usepackage{amsmath, graphicx, amssymb,tabls}
\usepackage{graphicx}
\usepackage{mathptmx}
\usepackage{amsmath,amssymb, amsthm}
\usepackage{rotating}

\usepackage{psfrag}
\usepackage[usenames]{color}

\renewcommand{\bbox}[1]{\mbox{{\pmb{$#1$}}}}
\newcommand{\beq}{\begin{eqnarray}}
\newcommand{\eeq}{\end{eqnarray}}
\newcommand{\beqn}{\begin{equation}}
\newcommand{\eeqn}{\end{equation}}
\newcommand{\bal}[1]{\begin{align} #1 \end{align}}

\newcommand{\expt}[1]{\langle #1 \rangle }
\newcommand{\Expt}[1]{\left\langle #1 \right\rangle }

\newcommand{\ra}{\rightarrow}

\newcommand{\nn}{\nonumber}

\begin{document}

\title{Linkage disequlibrium under recurrent bottlenecks}

\author{E. Schaper$^{a,}$\protect\footnote{Present address: 
\emph{\small Institute of Computational Science, ETH, CH-8092 Z\"urich, Switzerland} 
and \emph{\small Swiss Institute of Bioinformatics, CH-1015 Lausanne, Switzerland}}, 
A. Eriksson$^{b}$, M.Rafajlovic$^{a}$, S. Sagitov$^{c}$, and  B. Mehlig$^{a}$\\
$^a$\emph{\small Department of Physics, University of Gothenburg, SE-41296 Gothenburg, Sweden}\\
$^b$\emph{\small Department of Zoology, University of Cambridge, Cambridge, UK}\\
$^c$\emph{\small Mathematical Sciences, Chalmers University of Technology and University of Gothenburg, SE-41296 Gothenburg, Sweden}\\}

\begin{abstract}
Understanding patterns of selectively neutral genetic variation is essential in order to model deviations from neutrality, 
caused for example by different forms of selection. Best understood is neutral genetic variation at  a single locus,
but additional insights can be gained by investigating genetic variation at multiple loci. The corresponding patterns of
variation reflect linkage disequilibrium and provide information about the underlying multi-locus
gene genealogies. The statistical properties of two-locus genealogies have been intensively studied
for populations of constant census size, as well as for simple demographic histories such as exponential population growth, 
and single bottlenecks. By contrast, the combined effect of recombination and sustained demographic fluctuations 
is poorly understood.  Addressing this issue, we study a two-locus Wright-Fisher model of a population subject to recurrent bottlenecks.
We derive coalescent approximations for the covariance of the times 
to the most recent common ancestor at two loci. We find, first,  that an effective  population-size approximation describes
the numerically observed linkage disequilibrium provided that recombination occurs either much faster or much more slowly 
than the population size changes. Second, when recombination occurs frequently between bottlenecks but rarely within 
bottlenecks, we observe long-range linkage disequilibrium. Third, we show that in the latter case, 
a commonly used measure of linkage disequilibrium, $\sigma^2_d$ (closely related to $\hat r^2$), fails to capture long-range linkage
disequilibrium because constituent terms, each reflecting long-range linkage disequilibrium, cancel.
Fourth, we analyse a limiting case in which long-range linkage disequilibrium can be described
in terms of a Xi-coalescent process allowing for simultaneous multiple mergers of ancestral lines.
\end{abstract}
\maketitle


\section{Introduction}

 Genetic variation at single neutral loci has been investigated in great detail for population models under different
demographic processes, such as population expansions, single bottlenecks, or genetic hitchhiking caused by nearby selective sweeps
 (see for example \citep{Eri08} for a review of such models). All of of these models, either with non-overlapping generations, such as the Wright-Fisher model \citep{fis30:gen,wri31:evo}, or with continuous reproduction and mortality, such as the Moran model \cite{moran58}, have in common the underlying assumption of random mating.

Real biological populations 
exhibit abundance fluctuations on both short and long time scales, caused by e.\,g.\ environmental and ecological changes. Such size fluctuations in the form of repeated bottlenecks are characteristic of populations expanding into new territories: examples include the human out-of-Africa scenario \citep{LiuPrugnolle:2006, RamachandranDeshpande:2005}, the accompanying expansion of the parasite \emph{Plasmodicum falciparum} causing severe malaria \citep{TanabeMita:2010}, and the recolonization by the marine snail \emph{Littorina saxatilis} of  Sweden's west coast archipelago \citep{johannesson2003ele}. It is common practice to accommodate such fluctuations in the theory by using an effective population size instead of the census population size (see \citep{ewe82:neff} for a review of different measures of the effective population size). 

 Recent research has highlighted the importance of two competing time scales in the context of effective population-size approximations: the coalescent time scale (which is inversely proportional to the population size and reflects the time to the most recent common ancestor (MRCA) in populations with constant population size), and the time scale of the population-size fluctuations.
 Clearly, relatively slow demographic fluctuations can be ignored and the effective population size can be approximated by the initial population size \cite{Sjodin:2005}.
 In the opposite case of very frequent demographic fluctuations, it has been argued \citep{Wright:1938, cro70:int} that genetic variation of a population with varying population size is well described in terms of a population with constant population size $N_{\rm eff}$, given by the harmonic average of the population size $N_\tau$ in generation $\tau$:
\begin{equation}\label{eq:neff}
	N_{\rm eff}= \lim_{T\ra\infty}\Bigl(\frac{1}{T}\sum_{\tau=0}^T\frac{1}{N_\tau}\Bigr)^{-1}\,\,.
\end{equation}
See \citep{Sjodin:2005,JagersSagitov:2004,Wak09} for recent developments of this concept. 

Thus, for both fast and slow demographic fluctuations, 
the statistical properties of gene genealogies agree with those of
the  constant population-size model. By contrast, when both time scales are of the same order, it has been shown \citep{Nordborg:2003,Kaj:2003,Sjodin:2005,Eri10} that the distribution of total branch lengths in samples of one-locus gene genealogies does not in general agree with that predicted by the standard coalescent approximation. Especially, when subject to periodic fluctuations, the total branch length of gene genealogies may exhibit a maximum for periods matching the coalescent time scale \citep{Eri10}.
In summary, the effect of population-size fluctuations upon genetic variation of a single locus is well understood. 

But how do population-size fluctuations affect multi-locus patterns of genetic variation on the same chromosome?  Such patterns 
are influenced by recombination. Genetic recombination introduces a new time scale which is inversely proportional to the probability of recombination $r$ between a pair of loci in a single generation.

Genetic recombination plays an important role in shaping empirically observed multi-locus patterns of genetic variation in biological populations. Measures of linkage disequilibrium (LD) quantify the degree of association of genetic variation at pairs of loci on the same chromosome. Common measures of LD, such as $\hat{r}^2$ \cite{Hill:1968}, and its approximation $\sigma^2_d$ \cite{OhtaKim:1971,mcvean02}, depend upon the frequencies of alleles at two loci. These measures are thus closely 
related to the covariance of the times (i.\,e.\ the number of generations) to the MRCA
of the underlying gene genealogies \citep{mcvean02}.

In order to illustrate this concept, we show in Fig.~\ref{fig:examples} our results of the Wright-Fisher dynamics (grey lines) of a population experiencing recurrent bottlenecks, with random durations and separations between bottlenecks. Panels {\bf a} and {\bf b} show results for two different sets of parameters. 
Details are given in Section \ref{sec:model}. 
The single-locus properties of a similar model were studied by \citet{Sjodin:2005} and also by \citet{Eri10}. 
Each grey line in Fig.~\ref{fig:examples} shows the covariance of the times to the MRCA for two chromosomes at two loci, as a function of genetic distance along the chromosomes, corresponding to a single realisation of the sequence of bottlenecks, $\mathcal{D}$ (averaged over all pairs of loci the same distance apart). The red lines are the averages of the covariances within each panel. In panel {\bf a}, the bottlenecks happen frequently and have a short duration; in this case, the single-locus properties are expected to be in good agreement with those of a population with the constant effective population size, given by the harmonic time average, Eq.~(\ref{eq:neff}), of the population size $N_\tau$. For such populations, the coalescent approximation predicts \citep{griffiths81,hud83:pro,hud90:gen}
\begin{equation}\label{eq:const_pop_size}
	\text{cov}[t_{a(ij)},t_{b(ij)}] =x_{\rm eff}^2 \frac{R x_{\rm eff} + 18}{(Rx_{\rm eff})^2 + 13 R x_{\rm eff} + 18},
\end{equation}
where $t_{a(ij)}$ and $t_{b(ij)}$ are the times to the MRCA of two loci (called $a$ and $b$), in a sample of 
two chromosomes (denoted by $i$ and $j$). Moreover, $R = 2N_0r$ is the scaled recombination rate, and $x_{\rm eff}=N_{\rm eff}/N_0$ is the effective population size relative to $N_0$, the population size at the present time. Units of
time are chosen so that $\tau=\lfloor t N_0\rfloor$, and $\lfloor t N_0\rfloor$ is the largest integer not larger than $t N_0$. 
Effective population-size approximations, according to Eq.~({\ref{eq:const_pop_size}), are shown as dashed lines in Fig.~\ref{fig:examples}. In Fig.~\ref{fig:examples}{\bf a} we observe good agreement between Eq.~(\ref{eq:const_pop_size}) and the average covariance of the times to the MRCA obtained numerically. However, in Fig.~\ref{fig:examples}{\bf b}, Eq.~(\ref{eq:const_pop_size}) agrees with our numerically obtained covariance only for short genetic distances. For large genetic distances, by contrast, the numerical results show that the covariance decreases much more slowly than expected according to Eq.~(\ref{eq:const_pop_size}). Thus, the results shown in Fig.~\ref{fig:examples}{\bf b} imply long-range LD in two-locus gene genealogies. 

The examples shown in Fig.~\ref{fig:examples} open many questions in relation to multi-locus gene-genealogies. What are the conditions for the effective population-size approximation to be valid in the multi-locus case? Why does it fail when these conditions are not met? How significant are deviations of the exact result from the effective population-size approximation? Why does long-range linkage appear in some cases? How large are fluctuations around the covariance of the coalescent times, averaged over an ensemble of gene-genealogies and over different demographic histories? What is the significance of the fluctuations around such averages for data analysis? 

The aim of this paper is to provide answers to the above questions, by computing  the covariance of the times to the MRCA under a model of recurrent bottlenecks introduced in Section \ref{sec:model}. Our analysis enables us to qualitatively and quantitatively determine the effects of fluctuating population size on the two-locus statistics in terms of the time scales of population-size fluctuations, of coalescence, and of recombination. Using both a numerical and an analytical approach, we estimate the range of validity of the coalescent effective population-size approximation for the two-locus case. We find that the coalescent effective population-size approximation inevitably fails for large recombination rates: the failure is sometimes minor (as in the case shown in Fig.~\ref{fig:examples}{\bf a}) and sometimes significant (as in the case shown in Fig.~\ref{fig:examples}{\bf b}). By taking different limits of the parameters of the model, we provide both qualitative and quantitative understanding of how the constant effective population-size approximation may fail. Finally, when bottlenecks are severe, we show that gene genealogies can be approximated by those of a constant-sized population with simultaneous multiple pairwise coalescent events (the so-called Xi-coalescent described
by \citet{Schw:00} and \citet{MohSag:01}).  Last but not least, we demonstrate that $\sigma^2_d$ is surprisingly little affected by the long-range covariances.

\section{Model}\label{sec:model}

We use a Wright-Fisher model of a population of chromosomes, to trace the ancestry of $L$ loci on a pair of chromosomes backwards in time, until the most recent common ancestor of each locus is found. In each generation 
$\tau$, a set of parents is chosen randomly from the previous generation. 
Genetic recombination of a pair of chromosomes occurs independently with probability $r$ between each adjacent pair of loci.

To investigate the role of population-size fluctuations, we consider a model of recurrent bottlenecks in which the population size can take one of two values, $N_0$ or $x N_0$. Here $x N_0$ is the population size in the bottlenecks, and $0<x<1$. The probability of changing the population size, going one generation back in time, depends on the current population size. The switching probability is $p$ and $q$ when the current population size is $N_0$ and $x N_0$, respectively. Hence, the 
expected durations of the high and the low population-size phases are $1/p$ and $1/q$ generations, respectively. The population size in the first generation is taken to be $N_0$. We note that for the one-locus case, such a model  has been investigated by \citet{Sjodin:2005} and also by \citet{Eri10}. 

Fig.~\ref{fig:model} illustrates population-size fluctuations in this recurrent bottleneck model (panels {\bf a} and {\bf b}) and examples of gene genealogies of two loci (called $a$ and $b$) in a sample of two chromosomes (panels {\bf c} and {\bf d}). In panels {\bf c} and {\bf d}, generations with low population size ($xN_0$) are marked with yellow, otherwise the population size is $N_0$. Each chromosome is represented by a pair of lines (red and blue lines correspond to 
loci $a$ and $b$, respectively). In the generations where a common ancestor is found for a pair of ancestral lines, or a recombination between two loci occurs, the chromosomes are represented by circles instead of lines. The MRCA of a locus is shown as a filled circle. In some cases recombination causes the ancestry of one locus to become associated with a chromosome that lacks direct descendants in the sample (grey circles). The ancestries of such segments of DNA 
are irrelevant to the gene genealogy of the sample, and these ancestral lines are not traced further.

\section{Covariance of the times to the MRCA for two individuals}\label{sec:results}

As Fig.~\ref{fig:examples} shows, the covariance of the times to the MRCA at two loci (called $a$ and $b$) depends on the particular sequence of bottlenecks in the history of the population. Let $\text{cov}_\mathcal{D}[t_{a(ij)},t_{b(ij)}]$ denote the covariance conditional on the demographic history $\mathcal{D}$. In the following, we average the conditional covariance over random demographic histories to obtain (see red lines in Fig.~\ref{fig:examples}):
\begin{align}\label{eq:av_cov}
	\langle \text{cov}_\mathcal{D}[ t_{a(ij)}, t_{b(ij)}] \rangle
	&= \langle  \langle t_{a(ij)} t_{b(ij)} | \mathcal{D} \rangle -\langle t_{a(ij)} | \mathcal{D} \rangle \langle t_{b(ij)} | \mathcal{D} \rangle \rangle \nn\\
	&= \langle  t_{a(ij)} t_{b(ij)}  \rangle - \langle\langle t_{a(ij)} | \mathcal{D} \rangle^2 \rangle\,\,,
\end{align}
where $\langle \dots | \mathcal{D}\rangle$ denotes the expectation conditional on the particular demographic history $\mathcal{D}$. In the second equality we have used that the expected times to the MRCA are the same for both loci. Note that the averaged conditional covariance is not the same as the unconditional covariance of the times to the MRCA for the full process (given by $\expt{t_{a(ij)} t_{b(ij)}} - \expt{t_{a(ij)}}^2$). 

We now derive an approximate expression for $\langle{\text{cov}}_{\mathcal{D}}[t_{a(ij)}, t_{b(ij)}]\rangle$ using the coalescent approximation \citep{kin82:coa}, valid in the limit of large population sizes. Thus, in the following calculations, we assume $N_0\gg 1$, and $xN_0\gg 1$. As is usual in coalescent calculations, it is convenient to scale the rates $p$ and $q$ 
in the Wright-Fisher model  by a suitable representative of the population size. 
Defining the rates $\lambda = p N_0$, and $\lambda_\text{B} = q x N_0$ allows us to express the
probability for two lines to coalesce between two consecutive bottlenecks as  $(1+\lambda)^{-1}$. Correspondingly, 
the probability for two lines to coalesce during a single bottleneck is given by $(1+\lambda_{\rm B})^{-1}$. The units of time are chosen so that $\tau=\lfloor t N_0\rfloor$,  we take $R = 2N_0r$ as the scaled recombination rate as mentioned in the introduction.
In these units we denote the population size at time $t$ by $N(t)$, indicating that in the limit of $N_0\ra\infty$, $t$ becomes a continuous variable. 

We find the following  expression for the term 
$\langle\langle t_{a(ij)} | \mathcal{D} \rangle \langle t_{b(ij)} | \mathcal{D} \rangle \rangle\equiv\langle\langle t_{a(ij)} | \mathcal{D} \rangle^2 \rangle$,
occurring in Eq.~(\ref{eq:av_cov}):
\begin{equation}\label{eq:ta_tb}
\langle\langle t_{a(ij)} | \mathcal{D} \rangle^2 \rangle=\frac{\lambda _{\rm B} (2 x \lambda +\lambda _{\rm B}+3)+x \lambda  (x \lambda +x+2)+2}{(\lambda
   +\lambda _{\rm B}+1) (\lambda +\lambda _{\rm B}+2)}\,\,.
   \end{equation}
Details of the derivation of this result are summarised in Appendix~\ref{sec:t2}.

In order to evaluate the remaining term in Eq.~(\ref{eq:av_cov}), $\langle  t_{a(ij)} t_{b(ij)}  \rangle$,
we adapt the method described by \citet{Eri04} for calculating the covariance of the times to the MRCA at two loci, to the present model of recurrent bottlenecks. As can be seen in Fig.~\ref{fig:model}{\bf c}, {\bf d}, there are only a small number of possible combinations of ancestral lines in gene genealogies of two loci for two chromosomes. 
Thus, we can write down a Markov process for how states of the ancestral lines change along the gene genealogy. The corresponding graph is shown in Fig.~\ref{fig:diag_full}, where the vertices represent states (combinations of ancestral lines), and the edges represent transitions between the states (the transition rates $w_{ji}$ from $i$ to $j$ are shown along the edges from 
$i$ to $j$). The vertices labeled by a prime correspond to states in a bottleneck.

Following \citet{Eri04} we note that the expectation $\langle  t_{a(ij)} t_{b(ij)}  \rangle$
is determined by a six-dimensional sub-graph of the graph shown in Fig.~\ref{fig:diag_full}, consisting of the vertices  $1,2,3,1',2'$, and $3'$. 
Let $\bf{M}$ be the corresponding  $6\times 6$ transition matrix. Its off-diagonal elements are 
given by the transition rates, $w_{ji}$, from state $i$ to state $j$. The diagonal elements $M_{ii}$ are equal to the negative sum of the rates of all edges leaving node $i$ in the graph, i.\,e.\ 
$M_{ii}=-\sum_{j\neq i}M_{ji}$. At the present time ($t=0$), the system is in state $1$ in the graph. This is represented by the vector 
\bal{\bbox{v}  = \begin{bmatrix} 
	1 &  0 & 0 &  0 & 0 &  0 
\end{bmatrix}^{\sf T},
}
where $\sf T$ denotes the transpose. It is convenient to combine the transition rates $w_{ji}$ into vectors
and matrices as follows:
\bal{
	\bbox{u}&= 
	\begin{bmatrix} 
		1 & 0 & 0 & x^{-1} & 0  & 0
	\end{bmatrix}, \nn\\
	\bf{Q} &= 
	\begin{bmatrix} 
		0 & 2 & 2 & 0 & 0  & 0 \nn\\
		0 & 0 & 0 & 0 & 2x^{-1} & 2x^{-1} 
	\end{bmatrix}, \\
	\bf{K} &= \begin{bmatrix} 
		-(\lambda + 1) & \lambda_\text{B}x^{-1} \\
		\lambda & -(\lambda_\text{B}+1)x^{-1}
	\end{bmatrix},\\
	{\bbox c}&=\begin{bmatrix}
	1 & x^{-1}
		\end{bmatrix}\,. \nn
}
In terms of these vectors and matrices we can write \cite{Eri04}
\bal{\label{eq:exp_ta_tb}
\expt{t_{a(ij)} t_{b(ij)}} = \int_0^\infty dt_1\, t_1^2\, {\bbox u}\, {\rm e}^{{\bf M}\, t_1} {\bbox v} + \int_0^\infty  dt_1 \int_{t_1}^\infty  dt_2\, t_1 t_2\,{\bbox c}\, {\rm e}^{{\bf K}\left(t_2 - t_1\right)} \, {\bf Q}\, {\rm e}^{{\bf M}\,t_1} {\bbox v},
}
where $t_1$ and $t_2$ are the first and second coalescent events, respectively (i.\,e.\ $\min(t_{a(ij)},t_{b(ij)})$ and $\max(t_{a(ij)},t_{b(ij)})$). The first term corresponds to a common MRCA of loci $a$ and $b$ (i.\,e.\ a transition from states $1$ or $1'$ to state $5$), and the second to different MRCA (transitions from states $2$ or $3$ to state $4$, or from states $2'$ or $3'$ to state $4'$, followed by a transition to state $5$). The matrix $\bf{K}$ describes the change in the 
population size due to recurrent bottlenecks, whereas the vector $\bbox c$ contains the coalescent rates in each population-size regime. 
Both $\bf{M}$ and $\bf{K}$ have negative real eigenvalues. Hence, the integrals in Eq.~(\ref{eq:exp_ta_tb}) can be evaluated in terms of matrix inverses \cite{Eri04}:
\bal{\label{eq:exp_ta_tb2}
	\expt{t_{a(ij)} t_{b(ij)}} = 
	2\, {\bbox u}\, (-{\bf M})^{-3}\, {\bbox v} +
	2\, {\bbox c}\, (-{\bf K})^{-2} \left\{
	 (-{\bf K})^{-1}\, {\bf Q} + {\bf Q}\, (-{\bf M})^{-1}
	\right\} (-{\bf M})^{-2}\, {\bbox v}.
} 
Combining Eqs.~(\ref{eq:ta_tb}) and (\ref{eq:exp_ta_tb2}) yields
\bal{\label{eq:cov34}
\langle\mbox{cov}_\mathcal {D}[t_{a(ij)},t_{b(ij)}]\rangle =\frac{R^3C_3+R^2C_2+R C_1+C_0}{R^4D_4+R^3D_3+R^2 D_2+RD_1+D_0}\,\,,
}
where the coefficients $C_i$ and $D_i$ are functions of parameters $x$, $\lambda$ and $\lambda_{\rm B}$ given in Appendix C. 

We now discuss three special limits of this result. First, when the time to the first bottleneck is much longer than the expected times to the MRCA for two chromosomes at two loci in the large population-size regime (i.\,e.\ $\lambda \ll 1$), the subsequent bottlenecks are irrelevant to the mean covariance. In this case, Eq.~(\ref{eq:cov34}) reduces to Eq.~(\ref{eq:const_pop_size}), the expression valid in the case of constant population size with $x_\text{eff} = 1$.

Second, assume that the population-size fluctuations are rapid compared to all other processes (this case is described by taking the limit $\lambda\ra\infty$ and $\lambda_{\rm B}\ra\infty$ in such a way that the ratio $\lambda/\lambda_{\rm B}$ is kept constant). Keeping only the leading order of $\lambda$ and $\lambda_\text{B}$ in the numerator and the denominator of (\ref{eq:cov34}) yields Eq.~(\ref{eq:const_pop_size}), with $x_\text{eff}= (x\lambda + \lambda_\text{B})/(\lambda + \lambda_\text{B})$.
This is the harmonic time average (\ref{eq:neff}) of $N_\tau/N_0$ in the recurrent bottleneck model.

Third, we consider the case of severe bottlenecks ($x\ll 1$). Because the expected duration of a bottleneck is given by 
$x\lambda_\text{B}^{-1}$, this regime implies that bottlenecks are typically of short duration. Such demographic histories can occur during range expansions, where small groups of animals repeatedly colonize new areas (examples of this kind are given in the introduction). This case can be treated analytically by taking the leading order of $x^{-1}$ in the numerator and denominator of Eq.~(\ref{eq:cov34}). We find:
\bal{\label{eq:cov_short_strong}
\langle\text{cov}_\mathcal{D}[t_{a(ij)},t_{b(ij)}]\rangle \approx \frac{R^2A_2+RA_1+A_0}{R^2B_2+RB_1+B_0},
}
where $A_i$ and $B_i$ are functions of parameters $\lambda$ and $\lambda_{\rm B}$ given in Appendix \ref{sec:rates}. Note that this function reaches a plateau for large values of $R$: 
\begin{align}
\label{eq:cov_mult}
\langle\mbox{cov}_\mathcal{D}[t_{a(ij)},t_{b(ij)}]\rangle &\approx\frac{A_2}{B_2}\\
= &\frac{2\lambda_{\rm B}(1+\lambda_{\rm B})\lambda}{(1+\lambda_{\rm B}+\lambda)(2+\lambda_{\rm B}+\lambda)(9(2+\lambda)+\lambda_{\rm B}(27+8\lambda+\lambda_{\rm B}(10+\lambda_{\rm B}+\lambda)))}\,.\nn
\end{align}
This expression implies long-range linkage disequilibrium since the right hand side does not depend upon $R$.

\section{Comparison of the coalescent calculations to the Wright-Fisher simulations}

In order to further illustrate the role of the relevant time scales of genetic drift and recombination in shaping LD, we compare the full coalescent result, Eq.~(\ref{eq:cov34}), 
and the different limiting cases considered in Section~\ref{sec:results}, to the average covariance calculated from the Wright-Fisher simulations. The comparisons are 
shown in Fig.~\ref{fig:sim_theory_comparison}. 

The parameters used in Fig.~\ref{fig:sim_theory_comparison}{\bf a} correspond to rapid population-size fluctuations (the second example described in the previous section). As can be seen in Fig.~\ref{fig:sim_theory_comparison}{\bf a}, 
the agreement between the numerical result (red line) and the approximation (\ref{eq:const_pop_size}) (dashed line) is good for a wide range of recombination rates. A small disagreement appears at large values of $R$, more precisely 
at $R\approx 100$. This discrepancy is expected, as at such large recombination rates, the population-size fluctuations are no longer rapid compared to the process of recombination. In summary, in the case of rapid population-size fluctuations, the results of the Wright-Fisher simulations are well approximated by Eq.~(\ref{eq:const_pop_size}) as long as genetic recombination does not occur too frequently.

Fig.~\ref{fig:sim_theory_comparison}{\bf b} shows results for parameters corresponding to severe bottlenecks (the third example described in the previous section). 
As we noted already in the introduction, this case exhibits long-range linkage disequilibrium which 
cannot be accurately described by the effective population-size approximation
(\ref{eq:const_pop_size}). We observe that the average covariance curve can be divided into three regions where the curve behaves qualitatively differently. 
First, for very small recombination rates, the covariance can be approximated using the harmonic average effective population size. This is expected since, in this region, the population-size fluctuations are fast compared to the process of recombination. Second, the constant effective population-size approximation
 breaks down when $R \approx \lambda = 10$. For recombination rates larger than this value, by contrast, recombination occurs frequently between bottlenecks but rarely within, and multiple coalescent events during bottlenecks become frequent. Since the bottlenecks are short in this regime, this may lead to long-range LD.
In this regime, the covariance is approximated by Eq.~(\ref{eq:cov_short_strong}). We observe that, for a large range of recombination rates, Eq.~(\ref{eq:cov_short_strong}) is in excellent agreement with both the full coalescent result, Eq.~(\ref{eq:cov34}), and the simulations. Third, we see that the agreement between Eq.~(\ref{eq:cov_short_strong}) and the full coalescent result breaks down  when recombination events can no longer be ignored in the bottlenecks (i.\,e.\ when $R$ is of the same order as the rate of leaving a bottleneck, $\lambda_\text{B}x^{-1}$, or larger). For still larger recombination rates, only the full coalescent result agrees with the Wright-Fisher simulations. The slight deviations
between the Wright-Fisher simulations and the coalescent result (\ref{eq:cov34}), visible in Fig.~\ref{fig:sim_theory_comparison}, are discussed in the concluding section.

\section{The effect of recurrent bottlenecks upon  $\sigma^2_d$}

In the previous sections we have shown how sustained population-size fluctuations in
the form of recurrent bottlenecks give rise to long-range linkage disequilibrium,
measured by the covariance (\ref{eq:av_cov}) of the times to the MRCA.
An important question is how such population-size fluctuations affect
more common measures of linkage disequilibrium such as, for example, $\hat r^2$, and its approximation, $\sigma^2_d$ \citep{OhtaKim:1971,mcvean02}.
In this section we discuss the effect of recurrent bottlenecks upon $\sigma^2_d$.
While the covariance of the times to the MRCA is obtained by comparing two chromosomes,
the measure $\sigma_d^2$ is computed in a large sample of chromosomes. In the following
we consider the limit of infinite sample size. In this limit, and in terms of the expectations and covariances conditional
on the demographic history $\mathcal D$,
Eq.~(9) in \cite{mcvean02} becomes: 
\bal{
	\sigma^2_d = 
	\Expt{
	\frac{\text{cov}_{\mathcal D}[t_{a(ij)},t_{b(ij)}]-2\text{cov}_{\mathcal D}[t_{a(ij)},t_{b(ik)}] + \text{cov}_{\mathcal D}[t_{a(ij)},t_{b(kl)}]}{
	\expt{t_{a(ij)}|\mathcal D}^2 + \text{cov}_{\mathcal D}[t_{a(ij)},t_{b(kl)}]}}\,.
}
As before, $a$ and $b$ denote two loci, and $i$, $j$, $k$, and $l$ refer to four different chromosomes in a large sample. The main properties of this measure 
are determined by how the numerator depends on the recombination rate \cite{mcvean02}. In order to simplify the analysis, we therefore focus on the expected value of the numerator, i.\,e.\
\[
	\expt{\text{cov}_{\mathcal D}[t_{a(ij)},t_{b(ij)}]}-2\expt{\text{cov}_{\mathcal D}[t_{a(ij)},t_{b(ik)}]} + 
	\expt{\text{cov}_{\mathcal D}[t_{a(ij)},t_{b(kl)}]}.
\]
The covariance $\expt{\text{cov}_{\mathcal D}[t_{a(ij)},t_{b(ij)}]}$ is given by Eq.~(\ref{eq:cov34}). The covariances $\expt{\text{cov}_{\mathcal D}[t_{a(ij)},t_{b(ik)}]}$ and $\expt{\text{cov}_{\mathcal D}[t_{a(ij)},t_{b(kl)}]}$ can be calculated in the same way as Eq.~(\ref{eq:cov34}) was obtained, but starting from different initial conditions \citep{mcvean02,Eri04}. In our Markov representation, this corresponds to taking ${\bbox v} = [0\ 1\ 0\ 0\ 0\ 0]^T$ and ${\bbox v} = [0\ 0\ 1\ 0\ 0\ 0]^T$, respectively, in Eq.~(\ref{eq:exp_ta_tb}).

Fig.~\ref{fig:covariances} shows the relation between the covariances  $\langle{\rm cov}_\mathcal D[t_{a(ij)},t_{b(ij)}]\rangle$ (blue lines), $\langle{\rm cov}_\mathcal D[t_{a(ij)},t_{b(ik)}]\rangle$ (red lines), and $\langle{\rm cov}_\mathcal D[t_{a(ij)},t_{b(kl)}]\rangle$ (green lines). As can be seen, in a region of low values of $R$ each covariance is well approximated by the corresponding constant effective population-size approximation, as explained in Section~\ref{sec:results}. Note that in the case shown in panel {\bf b}, all three covariances exhibit a plateau at the same level, in approximately the same range of $R$ values. Thus, the linear combination of covariances in the numerator of $\sigma^2_d$ cancel 
an enhancement relative to the effective population-size approximation. This is the reason why  the plateau, present in each covariance, does not show in the numerator of $\sigma^2_d$. In other words, the information about the sustained population-size fluctuations is not preserved in $\sigma_d^2$.

\section{Discussion}

The aim of this paper was to provide an understanding on how sustained random population-size fluctuations influence gene-history correlations. Our conclusions are based on both analytical and numerical calculations, which we find to agree well. 
Using the particular population-size model, depicted in Fig.~\ref{fig:model}{\bf a}, we have derived an exact result for the covariance of the times to the MRCA of two loci, Eq.~(\ref{eq:cov34}). We have discussed three particular limits of our result. 
First, if the expected times to the MRCA of two loci are less than the expected time to the most recent bottleneck, our model reduces to the constant population-size model with the effective population size equal to the population size at 
the present time.
Second, if the population size fluctuates much faster than the remaining two processes (coalescence and recombination), the coalescent effective population-size approximation again works well, but with the effective population size given by the harmonic average (\ref{eq:neff}).
These two cases are consistent with earlier findings of investigations of single-locus properties of populations which exhibit population-size variations \citep{Sjodin:2005,Eri10}.
In the third case, bottlenecks are severe (large difference in population sizes in and between bottlenecks) and the time between bottlenecks is assumed to be brief. In this case, the result of the simulations depends on the relation between the time scale of recombination and the time scale of population-size changes. When recombination is the slowest process (i.\,e.\ when the recombination rate is sufficiently small), 
the coalescent effective population-size approximation (with the effective population size given by the harmonic average (\ref{eq:neff})) is a good approximation (this is essentially the same as in the second case). Conversely, when recombination is the fastest process, the covariance is approximately $1/R$ (same as in the first case). Finally, when recombination is intermediate, slow enough that recombination events during bottlenecks are rare, but fast enough to decorrelate gene genealogies between bottlenecks, the covariance exhibits a plateau. 
The plateau corresponds to an enhanced covariance of the times to the MRCA, 
with respect to that expected in the effective constant population-size case.  
Thus, in this case, pairs of distant loci are expected 
to exhibit a high degree of linkage.

These conclusions rely 
on analysing covariances averaged over different demographic histories. 
This raises the question how typical such averages are. In other words, how large are the 
fluctuations around the average? Fig. \ref{fig:examples} shows that 
in the case of severe bottlenecks, the fluctuations around the mean covariance are much higher than, for example, in 
the case of fast population-size fluctuations. 
This can be explained as follows: in the case of severe bottlenecks, the times to the MRCA of both loci 
are determined by the time to the bottleneck that hosts two pairwise mergers
(for very strong bottlenecks, this is simply the time to the first bottleneck).

The coalescent approximations employed in this paper assume large population sizes. While we generally find very good agreement between the coalescent approximations and the Wright-Fisher dynamics, we observe some deviations, in particular for severe bottlenecks and large recombination rates. The coalescent approximations assume that the time between two recombination events is exponentially distributed. However, this is only accurate in the limit of small recombination rates, $r$. Because the relevant parameter is $R = 2N_0r$, by increasing $N_0$ we reach a better agreement between the corresponding analytical and numerical results in the range of values of $R$ shown in Figs.~\ref{fig:examples} and~\ref{fig:sim_theory_comparison}. We have made additional simulations to confirm that this is the case (not shown). 

It is worth mentioning that the result  for the case of severe bottlenecks ($x\ll 1$), Eq.~(\ref{eq:cov_short_strong}),
can be understood in terms of the so-called Xi-coalescent approximation.  
Xi-coalescents form a broad  family of gene-genealogical models allowing for simultaneous multiple pairwise coalescent events (mergers). 
The Kingman coalescent is a special case, allowing only for pairwise mergers. 
See \citep{Schw:00,MohSag:01} for detailed descriptions of the family of Xi-coalescents. 
In the case of severe bottlenecks ($x\ll 1$), 
coalescent events during a bottleneck may appear as 
simultaneous multiple mergers (as pointed out  also by \citet{alea:coa}). 
We show in Appendix \ref{sec:rates} how the Markov process with simultaneous multiple mergers
is obtained in this case, and compute the corresponding transition rates $w_{ji}$ .
This provides an alternative way of deriving Eq.~(\ref{eq:cov_short_strong}). 
More importantly, it provides insight
into why the plateau forms. It turns out that the plateau arises as a direct consequence of simultaneous multiple mergers. 
We note that this means that long-range gene-history correlations
are also expected in other situations where simultaneous multiple mergers are important. 
Examples are populations with strongly skewed reproduction laws, or 
populations subject to selective sweeps.

We conclude with the observation that $\sigma^2_d$, a measure of LD, fails to show the plateaus present in its constituent covariances (this was observed already in \cite{Eri04} for the case of a single, recent bottleneck). Because of the close link between $\sigma^2_d$ and $\hat{r}^2$, a common measure of LD \cite{mcvean02}, this casts doubt on the suitability of such measures for characterising LD (another example is the measure $HR^2$ \citep{SabattiRisch:2002}), in populations that may have been subject to recent population bottlenecks and range expansions. A more accurate approach, especially for detecting long-range LD, may be to estimate the covariance of the times to the MRCA directly. For example, simulations show that the covariance of the number of mutations in small windows (e.\,g.\ a few hundred nucleotides long) can be used to estimate the covariance of the times to the
 MRCA \cite{Eri04}. However, it remains to investigate which observables are most suitable for detecting long-range dependencies in the underlying gene genealogies for more general demographic histories.


\textit{Acknowledgements}. Support by Swedish Research Council grants, 
the G\"oran Gustafsson stiftelse, and by the Centre for Theoretical 
Biology at the University of Gothenburg are gratefully acknowledged. AE was supported by a Philip Leverhume Award and a Biotechnology and Biological Sciences Research Council grant (BB/H005854/1).

\bibliographystyle{elsarticle-harv}
\bibliography{genetics}


\newpage
\appendix

\section{Calculation of $\expt{\expt{t_{a(ij)}|\mathcal{D}}^2}$}\label{sec:t2}

\newcommand{\xib}{\xi_\text{B}}
\newcommand{\xb}{y_\text{B}}
\newcommand{\lamb}{\lambda_\text{B}}
\newcommand{\etab}{\eta_\text{B}}

In this Appendix  we calculate $\expt{\expt{t_{a(ij)}|\mathcal{D}}^2}$ using a recursion. 
The resulting expression for $\expt{\expt{t_{a(ij)}|\mathcal{D}}^2}$ was used in evaluating
Eq.~(\ref{eq:av_cov}).  
In the Wright-Fisher model, let $\xi$ and $\xib$ be the expected times to the MRCA (i.\,e.\ $N_0\expt{t_{a(ij)}|\mathcal{D}}$, as in Section \ref{sec:results}) for bottleneck sequences starting outside and in the bottleneck, respectively. Furthermore, let $y = N_0^{-1}$ and $\xb = (xN_0)^{-1}$ (note that $y\xb^{-1}=x$). Furthermore, let $\eta$ and $\etab$ be independent stochastic variables which are unity with probability $p$ and $q$, respectively, and are zero otherwise. We have
\bal{\label{eq:recursion}
	\xi  &= 1 + (1-y )[(1-\eta)\xi' + \eta \xib'] \,\,,\nn\\
	\xib &= 1 + (1-\xb)[(1-\etab)\xib' + \etab \xi']\,\,,
}
where $\xi'$ and $\xib'$ have the same distribution as $\xi$ and $\xib$, respectively, but are statistically independent.
Taking the expected value of both sides, one obtains a linear system of equations for $\expt{\xi}$ and $\expt{\xib}$. Solving this system yields
\bal{
	\expt{\xi}  &= \frac{p (1 - y) + q(1 - \xb) + \xb}{ y \xb + p \xb (1 - y)  + q y (1 - \xb)}\,\,, \\
	\expt{\xib} &= \frac{p (1 - y) + q(1 - \xb) + y}{ y \xb + p \xb (1 - y)  + q y (1 - \xb)}\,\,.
}
In order to calculate $\expt{t_{a(ij)}|\mathcal{D}}^2$, we use  Eq.~(\ref{eq:recursion}) to write:
\bal{\label{eq:recursion2}
	\xi^2   &= 1 + 2(1-y)[(1-\eta)\xi' + \eta \xib'] + (1-y)^2(1-\eta)^2\xi'^2 +  (1-y)^2\eta^2\xib'^2\,\,, \nn\\
	\xib^2  &= 1 + 2(1-\xb)[(1-\etab)\xib' + \etab \xi'] + (1-\xb)^2(1-\etab)^2\xib'^2 +  (1-\xb)^2\etab^2\xi'^2\,\,.
}
Note that the terms containing $\xi' \xib'$ are absent because they contain factors $\eta (1-\eta)$ or $\etab(1-\etab)$, and $\eta$ and $\etab$ are either zero or unity. Taking the expected value of both sides of these equations, and using the 
property of independence, one again obtains a linear system that can be solved for $\expt{\xi^2}$ and $\expt{\xib^2}$:
\bal{
\expt{\xi^2} &=1+2(1-y)[(1-p)\expt{\xi}+p\expt{\xib}]+(1-y)^2(1-p)\expt{\xi^2}+(1-y)^2p\expt{\xib^2}\,\,,\nn\\
\expt{\xib^2}&=1+2(1-\xb)[(1-q)\expt{\xib}+q\expt{\xi}]+(1-\xb)^2(1-q)\expt{\xib^2}+(1-\xb)^2q\expt{\xi^2}\,\,.\label{eq:xi2_exp}
} 
 To leading order in $N_0^{-1}$, the solution to Eq.~(\ref{eq:xi2_exp}) is:
\bal{
\expt{\xi^2}
&=N_0^2\expt{\expt{t_{a(ij)}|\mathcal{D}}^2}
	=N_0^2 \frac{\lamb(\lamb+3  + 2 \lambda  x)+\lambda x(\lambda x+ x+2)+2}{
   \left(\lamb+\lambda +1\right) \left(\lamb+\lambda +2\right)}\,\,.\label{eq:exp_t2_wf}
}
Eq.~(\ref{eq:exp_t2_wf}) corresponds to the result (\ref{eq:ta_tb}) given in Section \ref{sec:results}.
Alternatively, Eq.~(\ref{eq:exp_t2_wf}) can be derived using  Eq.~(20) in \cite{Eri10}.

\section{Coefficients $C_i,D_i, A_i, B_i$ in formulae (\ref{eq:cov34}) and (\ref{eq:cov_short_strong})}\label{sec:coef}

In this Appendix we list the coefficients appearing in Eqs.~(\ref{eq:cov34}) and (\ref{eq:cov_short_strong}).
The coefficients appearing in Eq.~(\ref{eq:cov34})  are given by:
   \begin{align}
   C_0=&36 x^5\left(\lambda _{\rm B}+\lambda +3\right) \left(\lambda _{\rm B}+\lambda +6\right) (\lambda _{\rm B} (\lambda _{\rm B}
   \left(\lambda _{\rm B}+2 x \lambda +\lambda +4\right)\nn\\
   &+\lambda  (x ((x+2) \lambda +x+6)+1)+5)+x \lambda  (x (\lambda+1) (\lambda +3)+2)+2)\,\,,\nn\\
   C_1=&2 x^5 (\lambda _{\rm B} (\lambda _{\rm B} (\lambda _{\rm B} (\lambda _{\rm B} (\lambda _{\rm B}+3 x (\lambda +9)+2
   \lambda +13)+(3 x (x+2)+1) \lambda ^2\nn\\
   &+(x (55 x+76)+29) \lambda +252 x+59)+x^3 \lambda  (\lambda +1)(\lambda +27)\nn\\
   &+2 x^2 \lambda  (\lambda  (3 \lambda +58)+213)+x (\lambda  (\lambda  (3 \lambda +80)+370)+801)+2
   \lambda  (8 \lambda +55)+119)\nn\\
   &+x^3 \lambda  (\lambda +1) (\lambda +21) (2 \lambda +9)+x^2 \lambda  (\lambda 
   (\lambda  (3 \lambda +76)+452)+903)\nn\\
   &+x (\lambda  (\lambda  (31 \lambda +226)+647)+1044)+(3 \lambda +4) (5 \lambda+27))\nn\\
   &+x (\lambda  (x^2 (\lambda +1) (\lambda +3) (\lambda +6) (\lambda +15)+x (5 \lambda +18)(\lambda  (3 \lambda +16)+33)\nn\\
   &+44 \lambda +270)+468)+18 (\lambda +2))\,\,,\nn\\
   C_2=&x^5 (\lambda _{\rm B} (\lambda _{\rm B} (x \lambda _{\rm B} (3 \lambda _{\rm B}+2 x (4 \lambda +9)+4 (\lambda
   +7))+x (x^2 \lambda  (7 \lambda +39)\nn\\
   &+10 x (\lambda  (\lambda +7)+9)+\lambda  (\lambda +25)+89)+4\lambda )\nn\\
   &+x (2 x^3 \lambda  (\lambda +1) (\lambda +9)+x^2 \lambda  (\lambda  (8 \lambda +65)+129)\nn\\
   &+2 x(\lambda +2)^2 (\lambda +18)+\lambda  (9 \lambda +55)+116)+4 \lambda )\nn\\
   &+x (x (\lambda  (2
   x^2 (\lambda +1) (\lambda +3) (\lambda +6)+x (\lambda  (\lambda  (\lambda +22)+83)+102)\nn\\
   &+4 \lambda ^2+42 \lambda+96)+72)+26 (\lambda +2)))\,\,,  \nn\\
   C_3=&x^7 \left(\lambda _{\rm B}+\lambda +2\right) \left(\lambda _{\rm B} \left(\lambda _{\rm B}+2 x \lambda +3\right)+x \lambda  (x
   \lambda +x+2)+2\right) \,\,,
   \end{align}

   \begin{align}
  D_0=&36 x^5 \left(\lambda _{\rm B}+\lambda +1\right){}^2 \left(\lambda _{\rm B}+\lambda +2\right) \left(\lambda _{\rm B}+\lambda +3\right)
   \left(\lambda _{\rm B}+\lambda +6\right)\,\,,\nn\\
      D_1=&2 x^5 (\lambda _{\rm B}+\lambda +1) (\lambda _{\rm B}+\lambda +2) (\lambda _{\rm B} (\lambda _{\rm B} (13
   \lambda _{\rm B}+13 (x+2) \lambda +27 x+130)\nn\\
   &+13 (2 x+1) \lambda ^2+157 (x+1) \lambda +9 (19 x+39))\nn\\
   &+13 x(\lambda +1) (\lambda +3) (\lambda +6)+9 (\lambda +2) (3 \lambda +13))\,\,,\nn\\
         D_2=&x^5 (\lambda _{\rm B}+\lambda +1) (\lambda _{\rm B}+\lambda +2) (\lambda _{\rm B} (\lambda _{\rm B} (2
   \lambda _{\rm B}+4 x \lambda +39 x+2 \lambda +20)\nn\\
   &+x (2 x (\lambda  (\lambda +8)+9)+4 \lambda ^2+86 \lambda+247)+16 \lambda +54)\nn\\
   &+2 x^2 (\lambda +1) (\lambda +3) (\lambda +6)+13 x (\lambda +2) (3 \lambda +13)+18(\lambda +2))\,\,,\nn\\
           D_3=&x^6 \left(\lambda _{\rm B}+\lambda +1\right) \left(\lambda _{\rm B}+\lambda +2\right){}^2 \left(3 \lambda _{\rm B}+x (3 \lambda
   +13)+13\right)\,\,,\nn\\
           D_4=&x^7 \left(\lambda _{\rm B}+\lambda +1\right) \left(\lambda _{\rm B}+\lambda +2\right){}^2\,\,.
                      \end{align}

The coefficients appearing in Eq.~(\ref{eq:cov_short_strong})  are given by:
\begin{align}
A_0=&18 (\lambda _{\rm B}+1) \left(\lambda _{\rm B}+\lambda +3\right) \left(\lambda _{\rm B}+\lambda +6\right) \left(\lambda _{\rm B}
   \left(\lambda _{\rm B}+\lambda +3\right)+2\right)\,\,,\nn\\
A_1=&\left(\lambda _{\rm B}+1\right) \Bigl(\lambda _{\rm B} \bigl(\lambda _{\rm B} (\lambda _{\rm B}^2+2 (\lambda +6) \lambda _{\rm B}+\lambda 
   (\lambda +27)+47)+\lambda  (15 \lambda +83)+72\bigr)+18 (\lambda +2)\Bigr)\,\,,\nn\\
A_2=&2 \lambda_{\rm B} (\lambda_{\rm B}+1) \lambda  \,\,,  \nn\\
B_0=&18 \left(\lambda _{\rm B}+\lambda +1\right){}^2 \left(\lambda _{\rm B}+\lambda +2\right) \left(\lambda _{\rm B}+\lambda +3\right)
   \left(\lambda _{\rm B}+\lambda +6\right)\,\,,\nn\\
B_1=&\left(\lambda _{\rm B}+\lambda +1\right) \left(\lambda _{\rm B}+\lambda +2\right) \Bigl(\lambda _{\rm B} \bigl(13 \lambda _{\rm B}
   \left(\lambda _{\rm B}+2 (\lambda +5)\right)+\lambda  (13 \lambda +157)+351\bigr)+9 (\lambda +2) (3 \lambda +13)\Bigr)\,\,,\nn\\
B_2=&\left(\lambda _{\rm B}+\lambda +1\right) \left(\lambda _{\rm B}+\lambda +2\right) \left(\lambda _{\rm B} \left(\lambda _{\rm B} \left(\lambda
   _{\rm B}+\lambda +10\right)+8 \lambda +27\right)+9 (\lambda +2)\right)\,\,.
   \end{align}
\section{Severe bottlenecks: connection to the Xi-coalescent}\label{sec:rates}

In this Appendix, we turn our attention to a special case of population-size fluctuations: recurrent severe bottlenecks, that is $x\ra 0$. As we now show, single-locus gene genealogies in this limit are well approximated by the Xi-coalescent approximation (see also \cite{smre}).

Consider fixed values of 
$\lambda ~{\rm and}~\lambda_{\rm B}$ and take the limit of $x\ra 0$. Recall that the rate of leaving a bottleneck is given by $x^{-1}\lambda_{\rm B}$. Thus, as $x$ is being decreased, and $\lambda_{\rm B}$ is kept constant, the time between coalescent events hosted by a single bottleneck becomes shorter, ultimately leading to a failure of the Kingman coalescent approximation. In the limit of $x\ra 0$, multiple pairwise mergers in a bottleneck appear as a single simultaneous multiple merger. It turns out that one-locus gene genealogies in this case are well approximated by the Xi-coalescent approximation.

We show here that, in the case of severe bottlenecks, applying the Xi-coalescent approximation yields Eq.~(\ref{eq:cov_short_strong}) for the mean covariance $\langle\text{cov}_\mathcal{D}[t_{a(ij)},t_{b(ij)}]\rangle$.
Our method for calculating the term $\langle t_{a(ij)} t_{b(ij)}\rangle$ is described in the main text (see Eq.~(\ref{eq:exp_ta_tb}) in Section \ref{sec:results}). But in the Xi-coalescent approximation, described in this appendix, the Markov process is different than the one described in Section \ref{sec:results}. The corresponding graph is shown in Fig.~\ref{fig:diag_mult}. It consists of the same five states $1,\ldots,5$ shown in Fig.~\ref{fig:diag_full}, but in Fig.~\ref{fig:diag_mult} the states in bottlenecks are omitted, since the time system spends in a single bottleneck is short. The corresponding transition rates, $w_{ji}$, from state $i$ to $j$, are listed in Fig.~\ref{fig:diag_mult}.  In the following, we show how the rates $w_{ji}$ can be derived using the Xi-coalescent approximation. These rates determine the matrix $\bf M$, and vectors $\bbox u$, and $\bbox Q$, as well as $K$ and $c$, appearing in Eq.~(\ref{eq:exp_ta_tb}).

\subsection{Formulae for $w_{ji}$ under the Xi-coalescent approximation}\label{sec:rates1}
In a severe bottleneck, $l$ incoming lines are allowed to coalesce 
almost instantaneously into $b\le l-1$ outgoing lines.
Note that in Kingman's coalescent case one has $b=l-1$. 
In the Xi-coalescent, by contrast, $l$ lines are partitioned into $b$ families, such that $k_i$ families are of size $i=1,\ldots,l$ (see an illustration in Fig.~\ref{fig:part_lines}). By construction, the following condition must be satisfied:
\begin{equation}\label{eq:fam21}
l=\sum_{i=1}^l i k_i,~b=\sum_{i=1}^l k_i\,\,.
\end{equation} 
In our model, the collision rate $\phi_{\{l;k_1,\ldots,k_l\}}$ of $l$ lines colliding into a particular partition $\{l;k_1,\ldots,k_l\}$, such that Eq.~(\ref{eq:fam21}) is satisfied, 
is given by:
\begin{equation}\label{eq:phi1}
\phi_{\{l;k_1,\ldots,k_l\}}={ 1}_{\{b=l-1\}}+\lambda \Xi_{\{l;k_1,\ldots,k_l\}}
\end{equation}
where the first term stands for the Kingman coalescent, and the second term corresponds to the contribution from multiple mergers which occur at a rate $\lambda$. Given the probability, $C_{lb}$, that during a bottleneck $l$ lines collide into $b$ lines, $\Xi_{\{l;k_1,\ldots,k_l\}}$ can be calculated according to:
\begin{equation}
\Xi_{\{l;k_1,\ldots,k_l\}}=C_{lb}p_{\{l;k_1,\ldots,k_l\}}\nonumber\,\,.
\end{equation}
where  $p_{\{l;k_1,\ldots,k_l\}}$ is the probability of observing a particular partition $\{l;k_1,\ldots,k_l\}$ of $l$ lines. As shown by \citet{kin82:coa}, it is given by:
\begin{equation}
p_{\{l;k_1,\ldots,k_l\}}=\frac{(l-b)!b!(b-1)!}{l!(l-1)!}\prod_{i=1}^l (i!)^{k_i}\nonumber\,\,.
\end{equation}

The probability $C_{lb}$ is calculated as follows. Assume that $l$ lines enter a bottleneck. The probability that the bottleneck hosts $l-b$ coalescent events can be obtained in two steps. 
First, the total coalescent time needed to arrive from $l$ to $b$ lines should be less than or equal to the duration of the bottleneck. Second, the arrival at a state with $b-1$ lines must occur after the end of the bottleneck. For simplicity, we choose here to measure time in units of the population size during the bottleneck so that the coalescent rate is unity as in the standard coalescent case. In these units, the time in the bottleneck is exponentially distributed, with the parameter $\lambda_{\rm B}$. Accordingly, the following expression for $C_{lb}$ is obtained:
\begin{equation}\label{eq:Clb}
C_{lb}=\frac{\lambda_{\rm B}}{\binom{b}{2}+\lambda_{\rm B}}\prod_{i=b+1}^l\frac{\binom{i}{2}}{\binom{i}{2}+\lambda_{\rm B}} \nonumber\,\,.
\end{equation} 

The rate $\phi_{\{l;k_1,\ldots,k_l\}}$, given in Eq.~(\ref{eq:phi1}), is conditional on a particular partition. Thus the total collision rate of $l$ lines into any of partitions of type $\{l;k_1,\ldots,k_l\}$
is given by:
\begin{equation}\label{eq:rate0}
\phi_{\{l;k_1,\ldots,k_l\}}^{\rm tot}=\binom{l}{2}{ 1}_{\{b=l-1\}}+\lambda C_{lb} p_{\{l;k_1,\ldots,k_l\}}S_{\{l;k_1,\ldots,k_l\}}\,\,,
\end{equation}
where $S_{\{l;k_1,\ldots,k_l\}}$ denotes the number of possible ways of collisions of $l$ lines into a partition $\{l;k_1,\ldots,k_l\}$, such that restrictions in Eq.~(\ref{eq:fam21}) hold. It is given by \cite{sagitov03}:
\begin{equation}
S_{\{l;k_1,\ldots,k_l\}}=\frac{l!}{\prod_{i=1}^l (i!)^{k_i}k_i!}\nonumber\,\,.
\end{equation}
It follows from the latter expression that in the case of the standard coalescent, one has
$S_{\{l;l-2,1,0\ldots,0\}}=\binom{l}{2}$, as expected.

The graph corresponding to  the Markov process in the limit described in the beginning of this appendix, consists of five states $ 1,\ldots, 5$ (see Fig.~\ref{fig:diag_mult}). We now show  how the corresponding transition rates 
between states ${ 1,\ldots,5}$ can be derived from Eq.~(\ref{eq:rate0}). 

We observe that a collision of type ${\{2;0,1\}}$ describes a transition from either state $ 1$ or $ 4$, to $ 5$. It follows:
\begin{equation}\label{eq:rate1}
w_{51}=w_{54}=\phi_{\{2;0,1\}}^{\rm tot}\,\,.
\end{equation}
State $2$ consists of three chromosomal lines. A collision of a particular pair of lines, among the three lines, results in a transition from $2$ to $ 1$, while a collision of either of the two remaining pairs of lines results in a transition from $2$ to $ 4$. Because a collision of a pair of lines among three lines is of type $\{3;1,1,0\}$, we obtain the corresponding transition rates:
\begin{align}
&w_{12}=\frac{1}{3}\phi_{\{3;1,1,0\}}^{\rm tot}\,\,,\\
&w_{42}=\frac{2}{3}\phi_{\{3;1,1,0\}}^{\rm tot}\,\,.\end{align}
A collision of all three lines of state $2$ leads to a transition from state $2$ to $5$ at the rate:
\begin{equation}
w_{52}=\phi_{\{3;0,0,1\}}^{\rm tot}\,\,.
\end{equation}
Now consider transitions from state $3$. It consists of four ancestral lines. We analyse first a collision of a single pair of lines, that is a collision of type $\{4;2,1,0,0\}$. There are in total six different ways to pair lines: four choices describe a transition from state $3$ to $2$, and the remaining two lead to a transition from $3$ to $4$. Thus, we have:
\begin{align}
&w_{23}=\frac{2}{3}\phi_{\{4;2,1,0,0\}}^{\rm tot}\,\,.
\end{align}
Further, there are three possibilities for simultaneous collisions of two pairs of lines. Two possibilities result in a transition from $3$ to $ 1$, and one leads to a transition from $3$ to $5$ (see Fig.~\ref{fig:model}{\bf d}). It is also possible to obtain a collision of three lines, in which case the transition from $ 3$ to $4$ is observed. Further, a collision of all four lines results in a transition from $3$ to $5$. Thus, we obtain the following transition rates:
\begin{align}
&w_{13}=\frac{2}{3}\phi_{\{4;0,2,0,0\}}^{\rm tot}\,\,,\\
&w_{43}=\frac{1}{3}\phi_{\{4;2,1,0,0\}}^{\rm tot}+\phi_{\{4;1,0,1,0\}}^{\rm tot}\,\,, \\
&w_{53}=\frac{1}{3}\phi_{\{4;0,2,0,0\}}^{\rm tot}+\phi_{\{4;0,0,0,1\}}^{\rm tot}\,\,.\label{eq:rate2}
\end{align}
The remaining non-vanishing rates, $w_{21} =R$, and $w_{32}= R/2$, describe recombination transitions from state $1$ to $2$, and from $2$ to $3$. 

In Tab.~\ref{tab:1}, we summarise the formulae for $p_{\{l;k_1,\ldots,k_l\}}$, $S_{\{l;k_1,\ldots,k_l\}}$, $C_{lb}$, and
$\phi_{\{l;k_1,\ldots,k_l\}}^{\rm tot}$, for $l=2,3,4$ lines. Using the formulae in this table, the transition rates under the Xi-coalescent approximation can be calculated 
explicitly, in terms of the parameters $\lambda,\lambda_{\rm B}$, and $x$.

\subsection{Obtaining Eq. (\ref{eq:cov_short_strong}) 
under the Xi-coalescent approximation}
The transition rates $w_{ji}$, obtained in the previous
subsection, can be used for calculating 
$\langle t_{a(ij)} t_{b(ij)}\rangle$ according to the method explained in Section \ref{sec:results}. This calculation requires the elements of the $3\times 3$ matrix
${\bf M}$.
We find that the non-zero off-diagonal elements of  $\bf M$ are given by: 
\begin{align}
M_{12} &=
1+\lambda \lambda_{\rm B}\bigl((1+\lambda_{\rm B})(3+\lambda_{\rm B})\bigr)^{-1}\,\,,\nn\\
M_{13} & =
4 \lambda \lambda_{\rm B}\bigl((1 + \lambda_{\rm B}) (3 + \lambda_{\rm B}) (6 + \lambda_{\rm B})\bigr)^{-1}\,\,, \nn\\
M_{21} &=2 M_{32}= R\,\,,\nn\\
M_{23} & =
4 + 4 \lambda \lambda_{\rm B}\bigl((3 + \lambda_{\rm B}) (6 + \lambda_{\rm B})\bigr)^{-1}\,\,.\label{eq:M}
\end{align}
and the diagonal elements are given by:
\begin{equation}
M_{ii}=-\sum_{j\neq i} M_{ji},~{\rm for}~i=1,2,3\,\,.
\end{equation}
Further:
\begin{align}
	\bbox{u} &= 
	\begin{bmatrix} 
		&1+\lambda(1 + \lambda_{\rm B})^{-1} \\
		& 3 \lambda\bigl((1+\lambda_{\rm B})(3+\lambda_{\rm B})\bigr)^{-1} \\
		&2 \lambda (9 + \lambda_{\rm B})\bigl((1+ \lambda_{\rm B})(3+\lambda_{\rm B})(6+\lambda_{\rm B})\bigr)^{-1}
	\end{bmatrix}^{T}\,\,,\nn\\
	\bbox{Q} &= 
	\begin{bmatrix} 
		&0 \\
		& 2(1+\lambda \lambda_{\rm B}\bigl((1+\lambda_{\rm B})(3+\lambda_{\rm B})\bigr)^{-1})\\ 
		& 2 (1 + \lambda \lambda_{\rm B} (7 + \lambda_{\rm B})\bigl((1 + \lambda_{\rm B})(3 + \lambda_{\rm B})(6 + \lambda_{\rm B})\bigr)^{-1}) 
	\end{bmatrix}^{T}\,\,,\nn\\		
	K&=-1\,\,,\nn\\
	c&=1\,\,. \label{eq:c}
\end{align}
Note that $\bf M$, $\bbox u$, $\bbox Q$, $K$ and $c$ have dimensions different from those in Section \ref{sec:results}. The reason is that in the limit $x\ra 0$, the case described in this appendix, the states in bottlenecks are omitted. Combining Eq.~(\ref{eq:exp_ta_tb}) with Eqs.~(\ref{eq:M})-(\ref{eq:c}) yields Eq.~(\ref{eq:cov_short_strong}). This shows that the covariance of the times to the MRCA in the case of severe bottlenecks can be derived using the Xi-coalescent approximation.
\begin{figure}[t]
\psfrag{a}{\bf a}
\psfrag{b}{\bf b}
\psfrag{c}{\bf c}
\psfrag{d}{\bf d}
\psfrag{cov1}{\raisebox{+5mm}{\hspace*{-1cm}{$\mbox{cov}_\mathcal{D}[t_{a(ij)},t_{b(ij)}]$}}}
\psfrag{R}{$R$}
\centerline{\includegraphics[angle=0,width=15cm,clip]{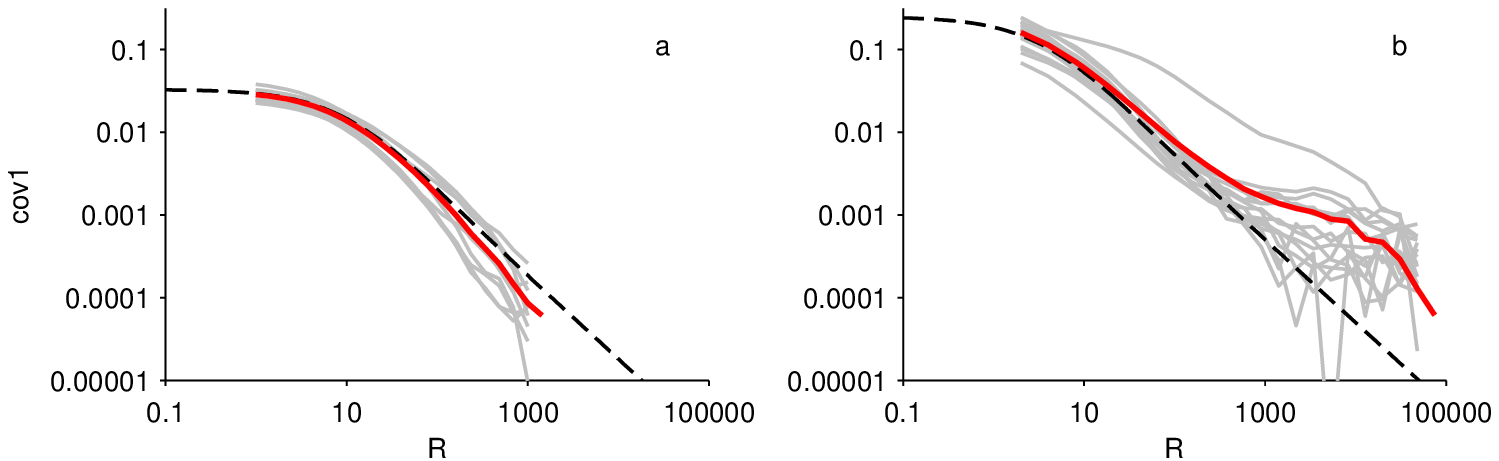}}
\caption{\label{fig:examples} 
The covariance of the times to the MRCA at two loci, in a sample of two chromosomes in a population subject to repeated bottlenecks (details in Section \ref{sec:model}). ({\bf a}) Rapid population-size fluctuations. Fisher-Wright simulations for ten random population-bottleneck sequences with $\lambda=100$, $x=0.1$, $\lambda_{\rm B}=10$, and $N_0=10^5$ (grey lines). Each grey line is obtained by first generating a random sequence of bottlenecks, and then averaging over an ensemble of $1000$ gene genealogies. The red line shows the covariance averaged over demographic histories. The dashed line shows the result of the effective population-size approximation, Eq.~(\ref{eq:const_pop_size}).
({\bf b}) Same, but for severe bottlenecks. Fisher-Wright simulations for fifteen randomly generated sequences of bottlenecks, with parameters $\lambda=10$, $x=5\cdot10^{-4}$, $\lambda_{\rm B}=10$, and $N_0=10^6$. Averaging is done over an ensemble of $100$ gene genealogies. 
}
\end{figure}
\begin{figure}[t]
\psfrag{a}{\raisebox{0mm}{\hspace*{+0cm}{\large \bf a}}}
\psfrag{b}{\raisebox{0mm}{\hspace*{+0cm}{\large \bf b}}}
\psfrag{c}{\raisebox{0mm}{\hspace*{0.1cm}{\large \bf c}}}
\psfrag{d}{\raisebox{0mm}{\hspace*{-0.2cm}{\large \bf d}}}

\psfrag{x1}{\raisebox{2mm}{\hspace*{-0.65cm}{\colorbox{white}{\large${\rm Exp}(\lambda)$}}}}
\psfrag{x2}{\raisebox{2mm}{\hspace*{-0.65cm}{\colorbox{white}{\large${\rm Exp}(\lambda_{\rm x})$}}}}
\psfrag{t}{\raisebox{0mm}{\hspace*{2.7cm}{\large $t$}}}
\psfrag{1}{\raisebox{0mm}{\hspace*{+0cm}{\large \bf 1}}}
\psfrag{2}{\raisebox{0mm}{\hspace*{+0cm}{\large \bf 2}}}
\psfrag{3}{\raisebox{0mm}{\hspace*{+0cm}{\large \bf 3}}}
\psfrag{4}{\raisebox{0mm}{\hspace*{+0cm}{\large \bf 4}}}
\psfrag{5}{\raisebox{0mm}{\hspace*{+0cm}{\large \bf 5}}}
\psfrag{N(t)}{\raisebox{-1mm}{\hspace*{-0.3cm}{\begin{sideways}\large$N(t)$\end{sideways}}}}
\psfrag{x N}{\raisebox{0mm}{\hspace*{-0cm}{\large$x N_0$}}}
\psfrag{N}{\raisebox{0mm}{\hspace*{-0.1cm}{\large$N_0$}}}
\psfrag{Locus a}{\raisebox{0mm}{\hspace*{0cm}{\large Locus $a$}}}
\psfrag{Locus b}{\raisebox{0mm}{\hspace*{0cm}{\large Locus $b$}}}
\psfrag{MRCA of locus a}{\raisebox{0mm}{\hspace*{0cm}{\large MRCA of locus $a$}}}
\psfrag{MRCA of locus b}{\raisebox{0mm}{\hspace*{0cm}{\large MRCA of locus $b$}}}
\psfrag{Chromosome}{\raisebox{0mm}{\hspace*{0cm}{\large Chromosome}}}
\psfrag{Ancestral line}{\raisebox{0mm}{\hspace*{0cm}{\large Ancestral line}}}
\psfrag{material not ancestral}{\raisebox{0mm}{\hspace*{0cm}{\large Material not ancestral}}}
\psfrag{to loci a and b}{\raisebox{0mm}{\hspace*{0cm}{\large to loci $a$ and $b$}}}
\psfrag{Period of low }{\raisebox{0mm}{\hspace*{0cm}{\large Period of low}}}
\psfrag{population-size}{\raisebox{0mm}{\hspace*{0cm}{\large population size}}}
\includegraphics[angle=0,width=18cm,clip]{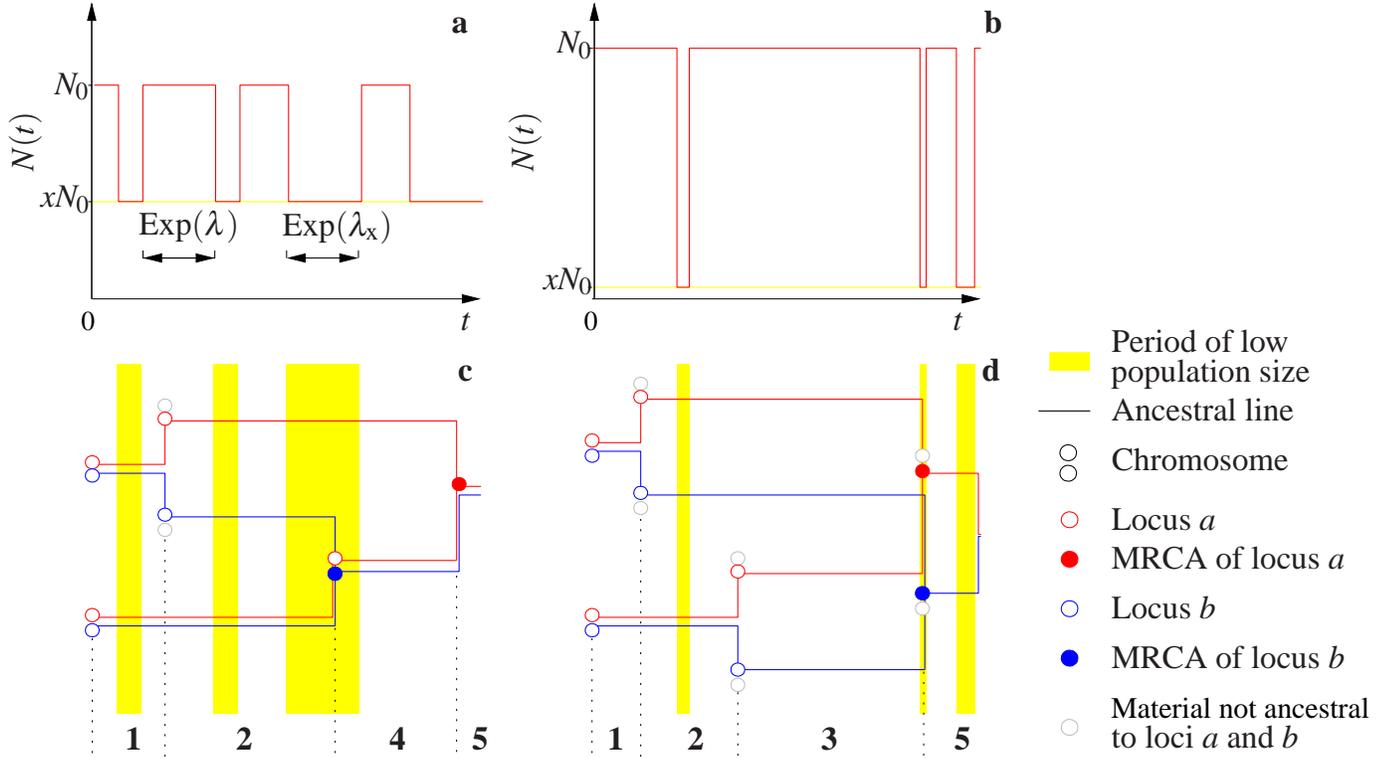}
\caption{\label{fig:model}Panels {\bf a} and {\bf b}  show two realisations
of the population-size curve, $N(t)$, backwards in time ($t=0$ denotes the present time). Initially, the population size is $N_0$. Going backwards in time, the population size randomly jumps between two values, $N_0$ and $x N_0$ ($x<1$), with the 
transition rates $\lambda$ (from $N_0$ to $x N_0$) and 
$\lambda_x\equiv x^{-1}\lambda_{\rm B}$ (from $x N_0$ to $N_0$). 
Panels {\bf c} and {\bf d} show schematically corresponding ancestral histories of two loci (blue and red empty circles correspond to two loci, called $a$ and $b$) subject to genetic recombination in a sample of two chromosomes. The yellow background depicts a time during which a population was subject to a bottleneck.
Two joint circles depict two loci in the same chromosome. States $1,\ldots,5$ denote the possible states of the system (they are explained in detail in Fig.~\ref{fig:diag_full}). Grey circles denote genetic material not ancestral to the sampled loci. Blue and red filled circles indicate that the corresponding loci have found their most recent common ancestor. Note that bottlenecks can host multiple coalescent events (mergers).  In the case of severe bottlenecks such multiple mergers appear as if instantaneous on the time scale of the gene genealogy. An example is shown in panel {\bf d}: an almost instantaneous  transition from state $3$ to state  $5$.
}
\end{figure}
\begin{figure}[t]

\begin{tabular}{@{}ll@{}}
\psfrag{24}{\raisebox{0mm}{\hspace*{+0.2cm}{\Large $1$}}}
\psfrag{25}{\raisebox{0mm}{\hspace*{+0cm}{\Large $1$}}}
\psfrag{26}{\raisebox{0mm}{\hspace*{0cm}{\Large $1$}}}
\psfrag{27}{\raisebox{0mm}{\hspace*{0cm}{\Large $R$}}}
\psfrag{28}{\raisebox{0mm}{\hspace*{+0cm}{\Large $4$}}}
\psfrag{6}{\raisebox{1mm}{\hspace*{-0.6cm}{\Large $R/2$}}}
\psfrag{7}{\raisebox{1mm}{\hspace*{0cm}{\Large $2$}}}
\psfrag{8}{\raisebox{0mm}{\hspace*{-0.2cm}{\Large $2$}}}
\psfrag{18}{\raisebox{0mm}{\hspace*{+0.2cm}{\Large $\lambda$}}}
\psfrag{21}{\raisebox{0mm}{\hspace*{+0cm}{\Large $\lambda$}}}
\psfrag{23}{\raisebox{1mm}{\hspace*{0cm}{\Large $\lambda$}}}
\psfrag{16}{\raisebox{1mm}{\hspace*{0.1cm}{\Large $\lambda$}}}
\psfrag{19}{\raisebox{0mm}{\hspace*{-0.1cm}{\Large $\lambda_{\rm x}$}}}
\psfrag{20}{\raisebox{1mm}{\hspace*{-0.3cm}{\Large $\lambda_{\rm x}$}}}
\psfrag{22}{\raisebox{1mm}{\hspace*{-0cm}{\Large $\lambda_{\rm x}$}}}
\psfrag{17}{\raisebox{1mm}{\hspace*{-0.1cm}{\Large $\lambda_{\rm x}$}}}
\psfrag{9}{\raisebox{0mm}{\hspace*{-0.2cm}{\Large $x^{-1}$}}}
\psfrag{10}{\raisebox{0mm}{\hspace*{+0cm}{\Large $x^{-1}$}}}
\psfrag{11}{\raisebox{1mm}{\hspace*{0cm}{\Large $x^{-1}$}}}
\psfrag{12}{\raisebox{1mm}{\hspace*{0cm}{\Large $R$}}}
\psfrag{13}{\raisebox{0mm}{\hspace*{+0cm}{\Large $4x^{-1}$}}}
\psfrag{14}{\raisebox{1mm}{\hspace*{+0cm}{\Large $R/2$}}}
\psfrag{15}{\raisebox{1mm}{\hspace*{0cm}{\Large $2x^{-1}$}}}
\psfrag{29}{\raisebox{1mm}{\hspace*{-0.6cm}{\Large $2x^{-1}$}}}

\includegraphics[angle=0,width=7cm,clip]{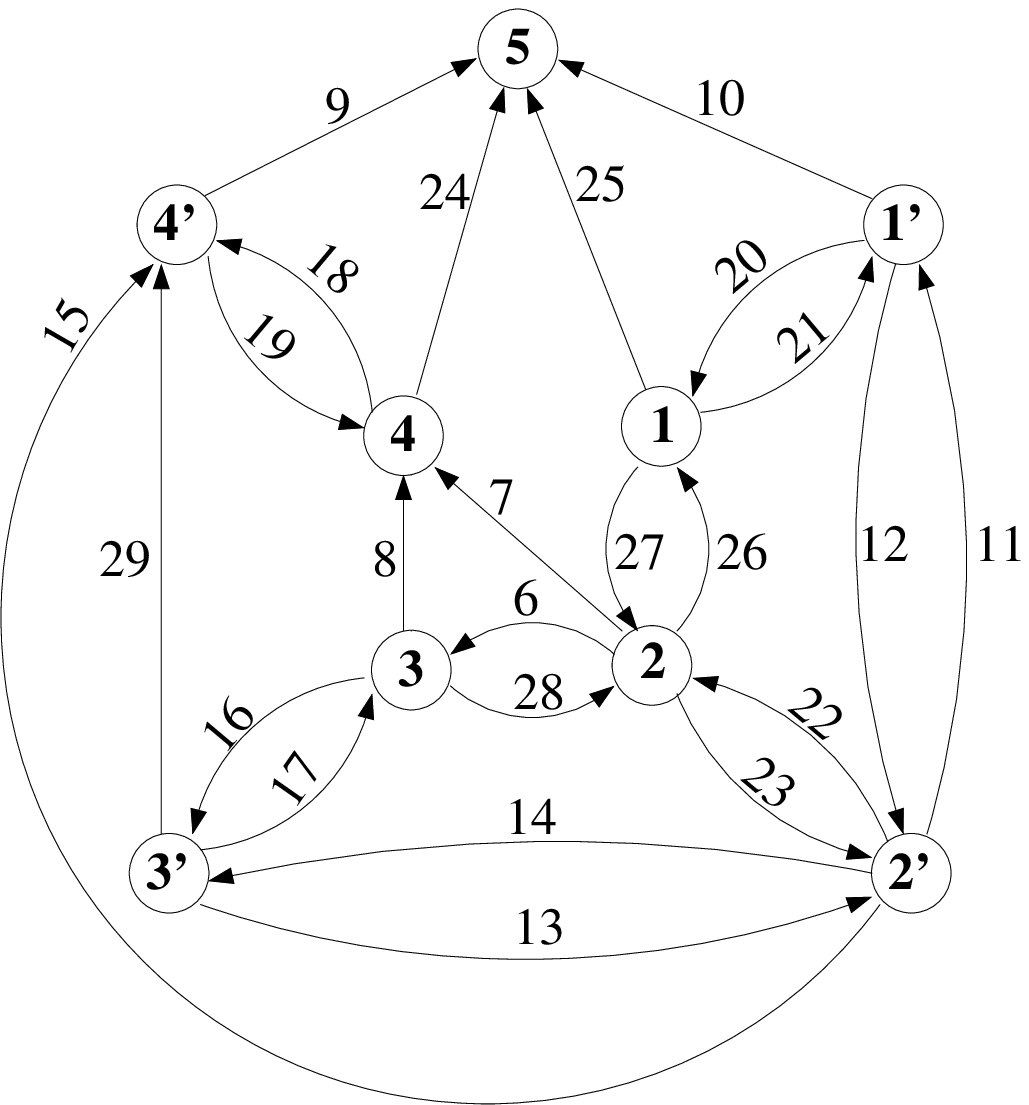}
&\hspace*{1cm}
   \raisebox{4cm}{
   \begin{tabular}{cl}
      \hline
      \large group $\ \ $ & \large state \\
      \hline
      \large\raisebox{0ex}{${\bf 1}, {\bf 1'}$} & \large $a_ib_i$,\,$a_jb_j$\\
                                  &   \large $a_ib_j$,\,$a_jb_i$\\
      \hline
      \large\raisebox{0ex}{${\bf 2}, {\bf 2'}$} & \large $a_i\circ$,\,$\circ b_i$,\,$a_jb_j$\\
                    &\large $a_ib_i$,\,$a_j\circ$,\,$\circ b_j$\\
                    & \large $a_i\circ$,\,$\circ b_j$,\,$a_jb_i$\\
                    & \large $a_ib_j$,\,$a_j\circ$,\,$\circ b_i$\\
      \hline
      \large {${\bf 3}, {\bf 3'}$}  & \large$a_i\circ$,\,$\circ b_i$,\,$a_j \circ$,\,$\circ b_j$ \\
      \hline
      \large\raisebox{0ex}{${\bf 4}, {\bf 4'}$} 
                   &\large  $a_i\bullet$,\,$a_j\bullet$\\
                   &\large  $\bullet b_i$,\,$\bullet b_j$\\
      \hline
      \large {${\bf 5}$} & \large$\bullet\bullet$ \\
      \hline
   \end{tabular}
   }
\end{tabular}

\caption{\label{fig:diag_full} Left: a graph showing the states and transition rates determining the ancestral history of two loci in a sample of two chromosomes, under the population model introduced in Section \ref{sec:model}. States where the population is in a bottleneck are marked with a prime. 
The final state is denoted by $5$ (in this state it does not matter whether the population is in a bottleneck or not). Arrows indicate transitions between states. The corresponding transition rates from state $i$ to $j$, $w_{ji}$, are displayed next to the lines. Note that $\lambda_{\rm x}\equiv \lambda_{\rm B}x^{-1}$. Right: a table of possible states of the system. Two loci considered are denoted by $a$ and $b$, and the corresponding chromosomes are indicated by $i$ and $j$. Empty circles denote genetic material not ancestral to sampled loci, and full circles denote the MRCA of a locus.
}
\end{figure}
\begin{figure}[t]
\psfrag{a}{\bf a}
\psfrag{b}{\bf b}
\psfrag{c}{\bf c}
\psfrag{d}{\bf d}
\psfrag{cov1}{\raisebox{+5mm}{\hspace*{-1cm}{$\langle\mbox{cov}_\mathcal{D}[t_{a(ij)},t_{b(ij)}]\rangle$}}}
\psfrag{R}{$R$}
\centerline{
\includegraphics[angle=0,width=15cm,clip]{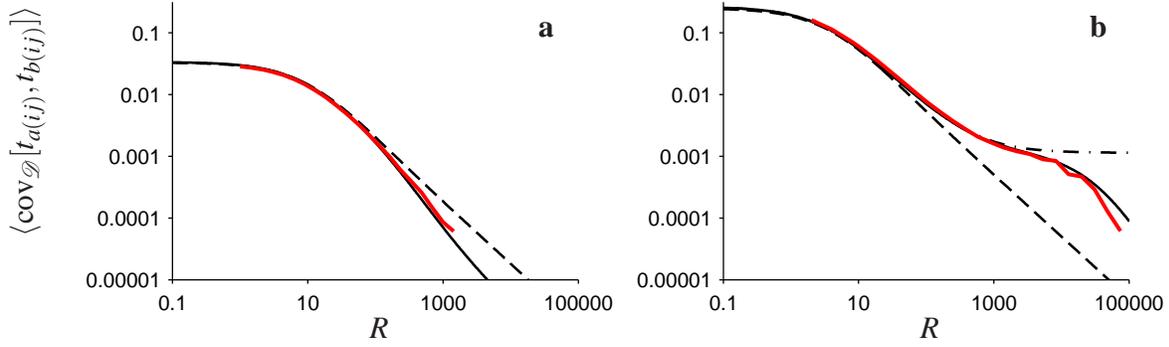}}
\caption{\label{fig:sim_theory_comparison} The 
covariance of the times to the MRCA of two loci averaged over random population-size histories. ({\bf a}) The red line shows the average covariance corresponding to $\lambda=100$, $x=0.1$, $\lambda_{\rm B}=10$, and $N_0=10^5$, determined numerically from Fisher-Wright simulations (same as in Fig.~\ref{fig:examples}{\bf a}).  The solid lines show our exact result, Eq.~(\ref{eq:cov34}), and the dashed lines show the coalescent effective population-size approximation, Eq.~(\ref{eq:const_pop_size}). The numerical result deviates from the effective population-size approximation when the recombination time-scale is the smallest ($R>100$). 
({\bf b}) The same as in {\bf a}, but for the short bottleneck case: $\lambda=10$, $x=5\cdot10^{-4}$, $\lambda_{\rm B}=10$, and $N_0=10^6$. The dashed-dotted line denotes the result of Eq.~(\ref{eq:cov_short_strong}), corresponding to the Xi-coalescent approximation. }
\end{figure}

\begin{figure}[t]
\psfrag{a}{\bf a}
\psfrag{b}{\bf b}
\psfrag{var(tij)}{\raisebox{5mm}{\hspace*{-0.6cm}{$\langle{\rm cov}_\mathcal D[t_a, t_b]\rangle$}}}
\psfrag{R}{$R$}
\psfrag{0.01}{\raisebox{0mm}{\hspace*{0cm}{\small$10^{-2}$}}}
\psfrag{1}{\small$10^0$}
\psfrag{100}{\raisebox{0mm}{\hspace*{-0cm}{\small$10^{2}$}}}
\psfrag{10000}{\raisebox{0mm}{\hspace*{-0cm}{\small$10^{4}$}}}
\psfrag{1e+06}{\raisebox{0mm}{\hspace*{-0cm}\small{$10^{6}$}}}
\psfrag{-6}{\raisebox{0mm}{\hspace*{-0.4cm}{\small$10^{-6}$}}}
\psfrag{-4}{\raisebox{0mm}{\hspace*{-0.4cm}{\small$10^{-4}$}}}
\psfrag{-2}{\raisebox{0mm}{\hspace*{-0.4cm}{\small$10^{-2}$}}}
\psfrag{0}{\raisebox{0mm}{\hspace*{-0.4cm}{\small$10^{0}$}}}
\centerline{\includegraphics[angle=0,width=15cm,clip]{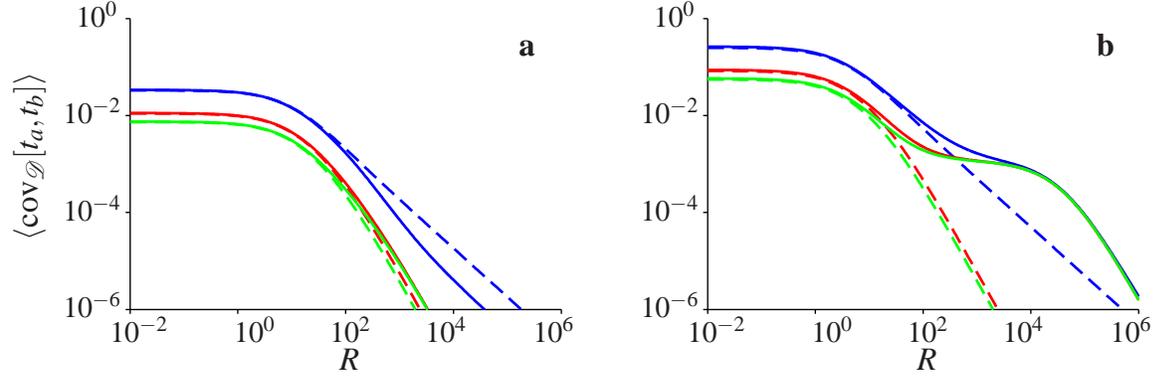}}

\caption{\label{fig:covariances} 
The relation between the covariances $\langle{\rm cov}_\mathcal D[t_{a(ij)},t_{b(ij)}]\rangle$ (blue lines), $\langle{\rm cov}_\mathcal D[t_{a(ij)},t_{b(ik)}]\rangle$ (red lines), and $\langle{\rm cov}_\mathcal D[t_{a(ij)},t_{b(kl)}]\rangle$ (green lines). In panels {\bf a} and {\bf b}, the values of the parameters $\lambda,\lambda_{\rm B}$, and $x$, are the same as in Fig.~\ref{fig:examples}{\bf a} and Fig.~\ref{fig:examples}{\bf b}, respectively. Exact results are shown as solid lines, whereas results obtained within the effective population-size approximation are shown as dashed lines. 
}
\end{figure}
\begin{figure}[t]
\includegraphics[angle=0,width=4cm,clip]{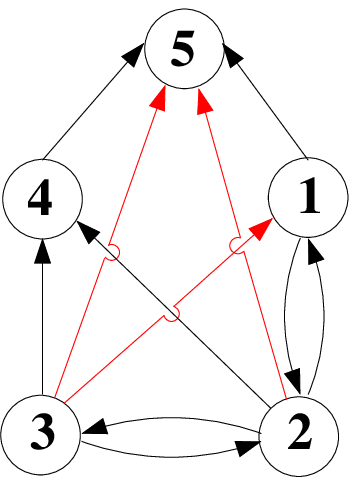}\hspace*{1cm}
\begin{minipage}[b]{2cm}
\begin{align*}
w_{12} &=
1+\lambda \lambda_{\rm B}\bigl((1+\lambda_{\rm B})(3+\lambda_{\rm B})\bigr)^{-1}\\
w_{13} & =
4 \lambda \lambda_{\rm B}\bigl((1 + \lambda_{\rm B}) (3 + \lambda_{\rm B}) (6 + \lambda_{\rm B})\bigr)^{-1} \\
w_{21} &=2 w_{32}= R \\
w_{23} & =
4(1 +  \lambda \lambda_{\rm B}\bigl((3 + \lambda_{\rm B}) (6 + \lambda_{\rm B})\bigr)^{-1}) \\
w_{42}& =2(1+\lambda \lambda_{\rm B}\bigl((1+\lambda_{\rm B})(3+\lambda_{\rm B})\bigr)^{-1})\\ 
w_{43}&= 2 (1 + \lambda \lambda_{\rm B} (7 + \lambda_{\rm B})\bigl((1 + \lambda_{\rm B})(3 + \lambda_{\rm B})(6 + \lambda_{\rm B})\bigr)^{-1}) \\
w_{51}&=w_{54}=1+\lambda(1 + \lambda_{\rm B})^{-1} \\
w_{52}&= 3 \lambda\bigl((1+\lambda_{\rm B})(3+\lambda_{\rm B})\bigr)^{-1} \\
w_{53}&=2 \lambda (9 + \lambda_{\rm B})\bigl((1+ \lambda_{\rm B})(3+\lambda_{\rm B})(6+\lambda_{\rm B})\bigr)^{-1}\\\end{align*}
\end{minipage}\hspace*{1cm}\
\caption{\label{fig:diag_mult} Left: a graph showing the states of the system in the limit $x\ra 0$, and possible transitions between them. States $1,\ldots,5$ are explained in Fig.~\ref{fig:diag_full}. Note that in this graph three colored arrows appear, denoting the simultaneous multiple mergers. They appear in this case because of the short time 
the system spends in a single bottleneck (see Fig.~\ref{fig:model}{\bf b}). By contrast, they are forbidden in the constant population-size case. Right: exact 
formulae for the transition rates, $w_{ji}$, from $i$ to $j$ (the corresponding entries of the matrix $\bf M$, and vectors $\bbox u$, and $\bbox Q$ are calculated using these rates) in terms of the parameters $\lambda$ and $\lambda_{\rm B}$.
}
\end{figure} 

\begin{figure}[t]
\psfrag{l1}{\raisebox{-3mm}{\hspace*{2cm}{\Large$l=4$}}}
\psfrag{l2}{\raisebox{-3mm}{\hspace*{-1cm}{\Large$b=1$}}}
\psfrag{a}{\raisebox{4mm}{\hspace*{0.2cm}{\Large {\bf a}}}}
\psfrag{b}{\raisebox{4mm}{\hspace*{0.2cm}{\Large{\bf b}}}}
\psfrag{present}{\raisebox{-0.2cm}{\Large{present}}}
\psfrag{t}{\raisebox{-0mm}{\hspace*{0cm}{\Large$t$}}}
\psfrag{N(t)}{\raisebox{6mm}{\hspace*{-0.2cm}{\Large $N(t)$}}}
\psfrag{N}{\raisebox{-0mm}{\hspace*{-0.5cm}{\Large{$N_0$}}}}
\psfrag{xN}{\raisebox{-0mm}{\hspace*{-0.5cm}{\Large{$xN_0$}}}}
\includegraphics[angle=0,width=10cm,clip]{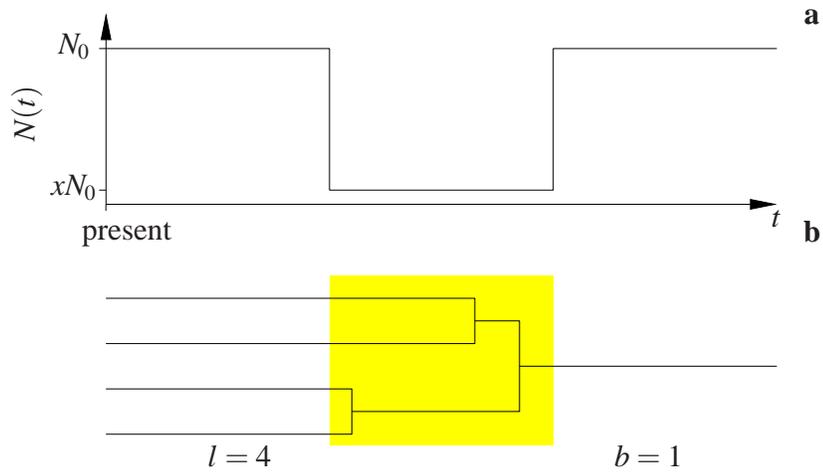}\hspace*{1cm}
\caption{\label{fig:part_lines} ({\bf a}) Realisation of a population-size history curve, $N(t)$. ({\bf b}) Partition of ancestral lines. Shown is a collision of $l=4$ lines (at the entrance of the bottleneck, indicated by the yellow background), into $b=1$ line. When the bottleneck is short, the three pairwise mergers during the bottleneck appear as a single instantaneous event.
}
\end{figure} 

{\renewcommand{\arraystretch}{2.5}
\begin{table}[tbhp]
\centering
\begin{tabular}{l l l l}
    \hline  \hline $C_{21}=\frac{1}{1+\lambda_{\rm B}}$ $ \ \ $   & $ p_{\{2;0,1\}}=1$ $\ \ $& $S_{\{2;0,1\}}=1$ $\ \ $& $\phi_{\{2;0,1\}}^{\rm tot}=1+\frac{\lambda}{1+\lambda_{\rm B}}$\\
      \hline $C_{31}=\frac{3}{(1+\lambda_{\rm B})(3+\lambda_{\rm B})}$ $ \ \ $  & $p_{\{3;0,0,1\}}=1$ $\ \ $ & $S_{\{3;0,0,1\}}=1$ $\ \ $ & $\phi_{\{3;0,0,1\}}^{\rm tot}=\frac{3\lambda}{(1+\lambda_{\rm B})(3+\lambda_{\rm B})}$\\
       \hline $C_{32}=\frac{3\lambda_{\rm B}}{(1+\lambda_{\rm B})(3+\lambda_{\rm B})}$ $ \ \ $  & $p_{\{3;1,1,0\}}=\frac{1}{3}$ $\ \ $ & $S_{\{3;1,1,0\}}=3$ $\ \ $ & $\phi_{\{3;1,1,0\}}^{\rm tot}=3+\frac{3\lambda \lambda_{\rm B}}{(1+\lambda_{\rm B})(3+\lambda_{\rm B})}$\\
       \hline  $C_{41}=\frac{18}{(1+\lambda_{\rm B})(3+\lambda_{\rm B})(6+\lambda_{\rm B})}$ $ \ \ $  & $p_{\{4;0,0,0,1\}}=1$ $\ \ $ & $S_{\{4;0,0,0,1\}}=1$ $\ \ $ & $\phi_{\{4;0,0,0,1\}}^{\rm tot}=\frac{18\lambda}{(1+\lambda_{\rm B})(3+\lambda_{\rm B})(6+\lambda_{\rm B})}$\\
        \hline $C_{42}=\frac{18\lambda_{\rm B}}{(1+\lambda_{\rm B})(3+\lambda_{\rm B})(6+\lambda_{\rm B})}$ $ \ \ $  & $p_{\{4;1,0,1,0\}}=\frac{1}{6}$ $\ \ $ & $S_{\{4;1,0,1,0\}}=4$ $\ \ $ & $\phi_{\{4;1,0,1,0\}}^{\rm tot}=\frac{12\lambda \lambda_{\rm B}}{(1+\lambda_{\rm B})(3+\lambda_{\rm B})(6+\lambda_{\rm B})}$\\ \cline{2-4}
          ~ $ \ \ $ & $p_{\{4;0,2,0,0\}}=\frac{1}{9}$ $\ \ $ & $S_{\{4;0,2,0,0\}}=3$ $\ \ $ & $\phi_{\{4;0,2,0,0\}}^{\rm tot}=\frac{6\lambda \lambda_{\rm B}}{(1+\lambda_{\rm B})(3+\lambda_{\rm B})(6+\lambda_{\rm B})}$\\        
          \hline  $C_{43}=\frac{6\lambda_{\rm B}}{(3+\lambda_{\rm B})(6+\lambda_{\rm B})}$ $ \ \ $  & $p_{\{4;2,1,0,0\}}=\frac{1}{6}$ $\ \ $ & $S_{\{4;2,1,0,0\}}=6$ $\ \ $ & $\phi_{\{4;2,1,0,0\}}^{\rm tot}=6+\frac{6\lambda \lambda_{\rm B}}{(3+\lambda_{\rm B})(6+\lambda_{\rm B})}$\\         
          \hline \hline        
          \end{tabular}
          \caption{\label{tab:1}Formulae necessary to  explicitly calculate the transition rates $w_{ji}$, listed in the graph in Fig.~\ref{fig:diag_mult}, according to Eqs.~(\ref{eq:rate1})-(\ref{eq:rate2}).}
\end{table}
}

\end{document}